\documentclass[conference]{IEEEtran}
\IEEEoverridecommandlockouts
\usepackage{cite}
\usepackage{amsmath,amssymb,amsfonts}
\usepackage{algorithmic}
\usepackage{graphicx}
\usepackage{textcomp}
\usepackage{xcolor}
\def\BibTeX{{\rm B\kern-.05em{\sc i\kern-.025em b}\kern-.08em
    T\kern-.1667em\lower.7ex\hbox{E}\kern-.125emX}}
\usepackage{amssymb}
\usepackage{mathrsfs}
\usepackage{stfloats}
\usepackage{dsfont}
\usepackage{setspace}
\usepackage{url}
\usepackage{hyperref}
\usepackage{tabularx}
\usepackage{multirow}
\usepackage{booktabs}
\usepackage{blindtext}
\usepackage{comment}

\newtheorem{definition}{Definition}
\begin{document}

\title{Practical Privacy-Preserving Data Science With Homomorphic Encryption: An Overview\\
{\footnotesize \thanks{The views and opinions expressed in this paper are those of the author and do not necessarily reflect the official policy or position of Banca d'Italia.}}
}
\author{\IEEEauthorblockN{Michela Iezzi}
\IEEEauthorblockA{\textit{ICT Department} \\
\textit{Bank of Italy}\\
Rome, Italy \\
michela.iezzi@bancaditalia.it}
}
\maketitle
\begin{abstract}
Privacy has gained a growing interest nowadays due to the increasing and unmanageable amount of produced confidential data. Concerns about the possibility of sharing data with third parties, to gain fruitful insights, beset enterprise environments; value not only resides in data but also in the intellectual property of algorithms and models that offer analysis results. This impasse locks both the availability of high-performance computing resources in the “as-a-service” paradigm and the exchange of knowledge with the scientific community in a collaborative view. Privacy-preserving data science enables the use of private data and algorithms without putting at risk their privacy. Conventional encryption schemes are not able to work on encrypted data without decrypting them first. Homomorphic Encryption (HE) is a form of encryption that allows the computation of encrypted data while preserving the features and the format of the plaintext. Against the background of interesting use cases for the Central Bank of Italy, this article focuses on how HE and data science can be leveraged for the design and development of privacy-preserving enterprise applications. We propose a survey of main Homomorphic Encryption techniques and recent advances in the conubium between data science and HE.
\end{abstract}
\begin{IEEEkeywords}
Privacy, encryption, data science, machine learning.
\end{IEEEkeywords}

\section{Introduction} \label{introduction}
Governments and institutions are currently at the forefront of the containment of COVID-19, and technology is one of the instruments that are put in place to counteract the pandemic: the term “Privacy-preserving contact tracing” has become popular and with it, concerns about the privacy of individuals are increasing. As citizens, we ask ourselves to what extent health officers use our tracing data, where and for how long they are stored, and how much it can be inferred from them \cite{Covid1}. 

As a central bank, we have access to multiple confidential data sources, structured and unstructured, due to the variegated nature of the public services we offer. Every single information asset has a different data owner, and inside a single data owner’s group, different levels of access exist, to control and prevent unauthorized processing and dissemination of data. In this scenario, the risk is a disempowerment of the benefits deriving from the potential collaboration between institutions and academic communities. Sharing data opens the door to the possibility of privacy breaches. Still, the other side of the coin is that paramount public results are obtainable only through a joint effort between the different owners of data silos and through fertilization with other knowledge domains. Collaboration provides a driver for innovation and competition. Additionally, other benefits, such as data storage and high-performance computation enabled by the cloud paradigm, could also be missed. It is clear that resorting to classical privacy preservation methods, such as Non Disclosure Agreement, is no longer a sufficient solution. 
To this purpose, in 2010, Anna Cavoukian \cite{Cavoukian} introduced the concept of Privacy by Design, intending to provide organizational and technical guidelines for the fulfillment of privacy requirements. The approach is holistic because it covers not only the technical aspects but also management and organizational processes. The main principle in Privacy by Design framework is that privacy requirements should be taken into account in the early stages of design and development of data application and should be maintained across all the application life cycle.

Privacy-preserving data science is a new emerging area, born from contamination between cryptography and data science. It can be described as a set of cryptographic techniques that aim at preserving the requirements of a particular definition of privacy of data and algorithms while allowing complex computations borrowed from statistics, machine learning, and natural language processing. In particular, this paper focuses on HE as a cryptographic technique for the design of private data science applications. The use of encryption on data exists already \emph{at-rest} and \emph{in-transit}, to avoid misuse or unauthorized access; however, to allow computation, data has to be decrypted, highlighting a potential privacy threat. HE realizes data encryption \emph{in-use}.

The demand for computing on encrypted data dates back to 1978 when Rivest, Adleman, and Dertouzos introduced the term \emph{privacy homomorphisms} related to the necessity of manipulating data banks \cite{RAD78}. They concluded the paper with two important open points that encouraged the scientific community to try to give them answer: (i) is it possible to apply privacy homomorphisms to real-world applications?, and (ii) there exists an algebraic system able to act as a ground for practical privacy homomorphisms? 

First HE schemes allowed limited computations on encrypted data, in terms of: (i) kind of available operations, i.e., only multiplications \cite{ElGamal} or only additions \cite{Paillier}; (ii) number of times the allowed operation can be performed or specified set of complex function to compute \cite{Acar}. In addition to known computational issues, the limitation in permitted operations posed an obstacle to real-world applications of HE.
It was only in 2009, with the introduction of Fully Homomorphic Encryption by Carl Gentry \cite{Gentry1} that the opportunity of performing a theoretically unlimited number of both additions and multiplications empowered research on the practical use of HE. Since then, the academic community spent much effort on optimizing HE schemes and developing libraries, and research activities recently extend the use of HE to data science applications, especially in the area of medicine \cite{BosMedical} and finance \cite{Cheon1}.

In this paper, we aim at investigating practical questions as: 
\begin{itemize}
    \item Is HE ready to fulfill privacy requirements in an enterprise context?
    \item which are the main available results in privacy-preserving data science with HE?
    \item which are the related challenges for real-world applications?
\end{itemize}
The contributions of the paper are summarized as follows: (i) we provide background on use cases of potential interest for Bank of Italy; (ii) we draw a state-of-art picture of main HE schemes and applications of HE to data science, touching on machine learning, natural language processing, and statistical analysis; (iii) discuss their potentialities and issues with particular regard to the integration of privacy-preserving computation in an enterprise context. The proposed overview is not exhaustive, as it is focused on exploring available results, having an eye to the Bank of Italy's industrial context.

The remainder of the paper is organized as follows. In Section II, we describe the enterprise background we are considering. In Section III, some essential preliminaries are given to posing the ground to Section IV, where fundamental theoretical results on HE schemes and available libraries to implement them are outlined. In Section V, an overview of the main results on privacy-preserving data science with HE is presented. In Section VI, the challenges of privacy-preserving data science in real-world applications are presented. Finally, Section VII concludes the paper.
\section{Motivation} \label{Motivation}

A central bank holds and employs different datasets to accomplish its institutional objectives. 
To name a few, we find datasets about Italian companies, datasets about loans granted and guarantees issued to households and firms, datasets about individuals subjected to revocation of payment cards because of missed payments, etc. \emph{Banking supervision}, support to \emph{oversight}, \emph{statistical research}, \emph{collateral eligibility} \cite{EDBT} are only some of data processing use cases based on the information sources mentioned above.

A compelling use case is counteracting money laundering, as argued in \cite{BellomariniUIF} for rule-based reasoning approach. The Italian Financial Intelligence Unit (FIU) operates within the Bank of Italy as an independent and autonomous body and receives about $100K$ Suspicious Transaction Reports (STRs) per year by financial intermediaries, such as banks, money transfers, dealers, etc. 
STRs contain extremely confidential data about the subject that executes the transaction, including his Personally Identifiable Information (PII), type, and amount of transaction. Furthermore, the intermediaries include a tentative manual classification of the type of suspicious phenomenon with a textual description.

Due to the confidential nature of data, only experts internal to the FIU, that are also the data owners, are allowed to access these reports, to prevent unwanted breaches that may pose at-risk intervention by investigative authorities. Such a timely classification of STRs might become unfeasible, taking into account the increasing yearly number of incoming reports. 
Here, the chance of a fruitful collaboration between data scientists and data owners arises, exploiting automatic classification of STRs enabled by machine learning algorithms. 
Though internal data scientists and external researchers are not fully trusted, they cannot access STRs, and data therein.
Currently, \emph{blind} access is obtained by an ad-hoc remote processing system, on a restricted tailored dataset with a fixed set of attributes only for a fixed set of selected years. Data owners and data scientists carefully discuss the choice of the set of attributes to obtain a trade-off between the efficacy of the classification and preservation of the privacy of data. This process may weaken the accuracy of automatic classification results because, e.g., feature selection (the selected attributes) is motivated by privacy reasons.

The mechanism of a remote processing system for blind access to confidential data is also employed for sharing statistical microdata of the Bank of Italy with internal and external researchers and other institutions. An example of statistical microdata shared for research purposes is the database that collects data from banks and financial companies (intermediaries) on the loans granted and guarantees issued to their customers (households and firms): the identification of the actors involved, financial intermediaries and physical persons, is prevented by the use of anonymization. Again, performing statistical analysis on anonymized data may lead to less accurate results; furthermore, the potential residual risk of re-identification is present \cite{BosMedical}.
\section{Preliminaries} \label{notation}

To present the main results from HE theory, we provide some basic definitions from maths and cryptography; the interested reader may refer to \cite{math1}, \cite{cryptoBoneh}. 

%
A \emph{group} $\mathcal{G}$ is a set with an operation $\diamond$ on its elements which: (i) is closed; (ii) has an identity; (iii) is associative; (iv) every element has an inverse. It is an \emph{abelian} group if it is commutative; it is denoted by $(\mathcal{G}, \diamond)$. 

A \emph{ring} $\mathcal{R}$ is a set with two operations $+$ and $\times$, respectively addition and multiplication, which is: (i) an abelian group with respect to the addition; (ii) closed under multiplication; (iv) associative with respect to multiplication; (v) has a multiplicative identity; (vi) multiplication and addition satisfy the distributive form. Furthermore, if multiplication is commutative, $\mathcal{R}$ is commutative. It is denoted by $(\mathcal{R},+,\times)$.

A \emph{group homomorphism} is a mapping $f:\mathcal{G} \rightarrow \mathcal{H}$ from a group $(\mathcal{G},\diamond)$ to another group $(\mathcal{H},\circ)$, such that the group operation is preserved, i.e.:
\begin{equation}
\label{eq_homomorphism}
f(g_1 \diamond g_2) = f(g_1) \circ f(g_2) 
\end{equation}
for all $g_1,g_2 \in \mathcal{G}$. The notion holds also for rings. 

%
A \emph{public key encryption scheme} $\varepsilon$ is a tuple of algorithms $(KeyGen_\varepsilon,Enc_\varepsilon,Dec_\varepsilon)$, where:
\begin{itemize}
\item $KeyGen_\varepsilon(\lambda)$ is the key generation algorithm with security parameter $\lambda$ as input, which outputs the pair $(pk,sk)$, public key and secret key, respectively;
\item $Enc_\varepsilon(pk,M)$ is the encryption algorithm, which takes the public key $pk$ and a set of plaintexts $M=(m_1,\ldots,m_n)$ from the ring of plaintexts $\mathcal{P}$ and outputs a set of cyphertext $\Psi=(c_1,\ldots,c_n)$ from the ring of cyphertext $\mathcal{X}$, i.e. $\Psi \leftarrow Enc_\varepsilon(pk,M)$, where $c_i \leftarrow Enc_\varepsilon(pk,m_i)$ for $i=1,\ldots,n$;
\item $Dec_\varepsilon(sk,\Psi)$ is the decryption algorithm, which takes the secret key $sk$ and a set of cyphertexts $\Psi$ and outputs a set of  plaintexts $M$; i.e. $M \leftarrow Dec_\varepsilon(sk,\Psi)$, where $m_i \leftarrow Dec_\varepsilon(sk, c_i)$.
\end{itemize}

A public key encryption scheme $\varepsilon$ is said to be \emph{correct} if:
\begin{equation}
\label{PKE_correct}
\begin{split}
\forall m_i \in \mathcal{P},~Pr[(pk,sk) \leftarrow KeyGen_\varepsilon(\lambda) : \\  Dec_\varepsilon(sk,Enc_\varepsilon(pk,m_i))=m_i] = 1
\end{split}
\end{equation}
The correctness property ensures that $Dec_\varepsilon(sk,c_i)$ is exactly the same $m_i$ in input to the encryption scheme $\varepsilon$.

A circuit $C$ is an acyclic directed graph, with $n$ inputs and $m$ outputs. Nodes are gates, which computes a function $f$. Edges are wires, arranged in a well defined order. In \emph{Boolean} circuits, inputs are booleans and gates are defined over a basis $\mathcal{B}$ of logic gates. $\mathcal{B}$ can be functionally complete: every boolean function $f:\{0,1\}^m \rightarrow \{0,1\}$ can be constructed starting from $\mathcal{B}$. An example of functionally complete basis is $\{AND,OR,NOT\}$.
Informally, the \emph{size} $S_f$ of a circuit $C$ is the number of gates, while the \emph{depth} $D_f$ is the length of its longest path from an input to the output\cite{Vollmer}.
\section{Homomorphic Encryption in a Nutshell} 
\label{HE_Theory}

Starting from (\ref{eq_homomorphism}), we can state that an encryption scheme is said to be \emph{homomorphic} over an operation $\diamond$ defined on ring of plaintexts $\mathcal{P}$ if \cite{HE_book1}:
\begin{equation} 
\label{HE_def1}
Enc_\varepsilon(pk,m_1 \diamond m_2) =  Enc_\varepsilon(pk,m_1) \circ Enc_\varepsilon(pk,m_2) 
\end{equation}
for $\circ$ defined on ring of cyphertexts $\mathcal{X}$. 

Furthermore, it is straightforward: 
\begin{equation} 
\label{HE_def2}
m_1 \diamond m_2 = Dec_\varepsilon(sk,c_1 \circ c_2)
\end{equation}

More generally, an \emph{HE} scheme $\varepsilon$ is a tuple of algorithms
:
\begin{equation*}
(KeyGen_\varepsilon(\lambda), Enc_\varepsilon(pk,M), Dec_\varepsilon(sk,\Psi), Eval_\varepsilon(pk,f,\Psi))
\end{equation*}
where $f \in \mathcal{F_\varepsilon}$ is belonging to the family of admissible function. 

For every $f$, we have that:
\begin{equation} 
\label{eq_eval}
   \overline{\Psi} \leftarrow Eval_\varepsilon(pk,f,\Psi) 
\end{equation}

In (\ref{eq_eval}), the algorithm $Eval_\varepsilon(pk,f,\Psi)$ picks the public key and performs the evaluation of the function $f$ on $\Psi$, returning a set of cyphertexts $\overline{\Psi}$.

As seen in Section \ref{notation}, the function $f$ may be seen as
a Boolean circuit $C^*$ on its input. Accordingly, $\mathcal{F_\varepsilon}$ maps to $\mathcal{C}$, the set of permitted circuits, i.e. the circuits that the HE scheme is able to evaluate; (\ref{eq_eval}) can be rewritten as:

\begin{equation}
\label{eq_eval_circuit}
    \overline{\Psi} \leftarrow Eval_\varepsilon(pk,C^*,\Psi) 
\end{equation}
The related circuit $C^*$ is evaluated over the input set $\Psi$ and returns the set $\overline{\Psi}$.
Let us introduce some desirable properties of our HE schemes \cite{ArmGuide}. First of all, a correct decryption of the encrypted result of the computation should be achieved. 
\begin{definition}
An HE scheme $\varepsilon$ is \emph{correct} if it holds \ref{PKE_correct} and:
\begin{equation}
\label{HE_complete_1}
Pr[Dec_\varepsilon(sk,\overline{\Psi})=C^*(M)] = 1-negl(\lambda)
\end{equation}
where $negl()$ is a negligible function of the security parameter.
\end{definition}

HE schemes should fulfill the following properties to avoid trivial cases.
\begin{definition}
An HE $\varepsilon$ is \emph{compact} if there exists a polynomial $g$ such that the size of $\overline{\Psi}$ is not more than $g(\lambda)$ bits and is independent of the length of the circuit.
\end{definition}

Compactness poses an upper bound for the length of the evaluated ciphertext: size of the ciphertext should be independent of the evaluated circuit, and it does not grow with its complexity.
\begin{definition}
An HE scheme is \emph{efficient} if there is a polynomial $s$ such that for every circuit of size $S_f$ the algorithm $Eval$ has complexity at most $S_fs(\lambda)$.
\end{definition}

It assures that decrypting the evaluated ciphertext and decrypting encrypting values takes roughly the same amount of time, i.e. polynomial in the security parameter. 
\begin{definition}
An HE scheme \emph{compactly evaluate} if it is correct and it is compact. 
\end{definition}

HE schemes are based on the hardness of known problems \cite{ArmGuide}, which also assures the security of such schemes. It is worth highlighting that the scheme's computational complexity also derives from such hardness, a crucial point in choosing HE schemes parameters, and, consequently, the security level, is guaranteeing a trade-off between privacy and practicability.

Using the classification in \cite{Acar}, we can distinguish three main categories of HE schemes according to the type of circuit to be evaluated and to the size and depth of the considered circuit.
Let us overview the principal families of HE schemes; for the sake of brevity, we do not provide an in-depth description of the underneath algorithms but give references for details.
\subsection{Partially Homomorphic Encryption (PHE)} 
\label{PHE}
PHE schemes allow only one type of operation, namely addition or multiplication.
A HE scheme is said to be \emph{homomorphically additive} if it is correct for the family of boolean circuits composed of only $XOR$ gates, or equivalently:
\begin{equation*}
Enc_\varepsilon(pk,m_1) * Enc_\varepsilon(pk,m_2) =  Enc_\varepsilon(pk,m_1 + m_2) 
\end{equation*}

A classic example of an additive PHE scheme is Paillier \cite{Paillier}, which relies on the Decisional Composite Residuosity Assumption (DCRA). Any multiplication on encrypted data corresponds to the addition of plaintexts. 

A homomorphic scheme is said to be \emph{homomorphically multiplicative} if it is correct for the family of Boolean circuits composed of only $AND$ gates, or equivalently: 
\begin{equation*}
Enc_\varepsilon(pk,m_1) * Enc_\varepsilon(pk,m_2) =  Enc_\varepsilon(pk,m_1 * m_2)
\end{equation*}

El Gamal \cite{ElGamal} proposed a multiplicative scheme, which is based on the discrete logarithm problem. Any multiplication on encrypted data is equal to the multiplication on plaintexts.
\subsection{Somewhat Homomorphic (SWHE) and Fully Homomorphic Encryption (FHE)} 
\label{SWHE_FHE}
The landmark result towards complex computation on encrypted data came in 2009 with Gentry’s work \cite{Gentry1}. Before then, SWHE schemes were available: these schemes are correct only for a specified set of the permitted circuit. Furthermore, in some schemes, the resulting cyphertext size grows in the evaluating process of AND and XOR gates, failing in fulfilling the compactness requirement. In some other cases, the scheme is not secure. Notable examples of such a line of research can be found in \cite{Acar} and references therein. 
After 2009, the FHE scheme introduced by Gentry led to a set of related developments for schemes that are not fully homomorphic but suitable for real-world applications. Before introducing the steps to obtain FHE, let us define what an FHE scheme is.
\begin{definition}
A homomorphic scheme is said to be \emph{fully homomorphic} if it is efficient and compactly evaluates for the set of all Boolean circuits.
\end{definition}

The main idea behind \cite{Gentry1} resides in constructing a FHE scheme starting by a SWHE scheme, and then applying two operations, \emph{squashing} and \emph{bootstrapping}. As a starting point for an informal discussion \cite{Gentry2}, let us introduce the symmetric version of a simple, but popular, SWHE scheme, known as DGHV \cite{DGHV}, that works as follows. For $m=\{0,1\}$, $p$ odd integer, $r$ and $q$ randomly chosen integers, it holds:
\begin{equation}
    Enc_\varepsilon(pk,m) : c \leftarrow m + pq + 2r
\end{equation}
\begin{equation}
    Dec_\varepsilon(sk,c) : m \leftarrow (c~\text{mod}~p)~\text{mod}~2
\end{equation}

DGHV is based on the Approximate Greatest Common Divisors (AGCD) problem: the problem of recovering $p$ from near multiples of $p$. When the random noise $r$ is smaller than $p$, we can correctly decrypt and recover the plaintext.

The scheme is both additively and multiplicatively homomorphic: 
\begin{equation}
    c_1 + c_2 \simeq m_1 + m_2 + 2(r_1+r_2)
\end{equation}
\begin{equation}
    c_1c_2 \simeq m_1m_2 + r_1r_2
\end{equation}
As we observe, with the evaluation of addition, noise has doubled, while with the evaluation of multiplication, noise grows quadratically. It derives that, in order for the SWHE scheme to be correct, it is feasible to perform only a limited number of operations. 
To overcome noise issue, Gentry introduced \emph{bootstrapping}, that works as follows \cite{Gentry1}. It performs decryption of the ciphertext before the noise exceeds a threshold: this removes the noise with respect to the first encryption key. Then, encryption is performed again using a new key. The new encryption key should be chosen carefully to obtain a less noise level than the removed one. This operation is repeated each time the noise grows over the permitted threshold, allowing the scheme to compute an unlimited number of operations. A requirement for the SWHE scheme to be bootstrappable is the homomorphic evaluation of its decryption circuit: this means that it is desirable to have the decryption circuit in the set of permitted circuits of the SWHE scheme.  
Since the latter requirement is not valid for a large class of SWHE schemes, it could be necessary to perform a \emph{squashing} operation \cite{Gentry1} to enable bootstrapping. For a SWHE scheme, squashing decreases the circuit depth of the decryption algorithm, allowing the scheme to handle it. However, squashing causes an increase in the cyphertext length and the introduction of additional hardness assumptions, such as the \emph{Sparse Subset Assumption} (SSP). We do not dive into further explanation, which goes beyond the purpose of this paper; more details about how squashing works can be found in \cite{Gentry1} and \cite{DGHV}. Leveled Fully HE (LHE) schemes are closely related to bootstrapping since they are in some sense SWHE scheme: the application of bootstrapping up to a fixed level $d$ results in a scheme that is efficient and compactly evaluates for all Boolean circuits of depth $d$ \cite{ArmGuide}. 

Gentry's results gave rise to a fruitful line of research on HE \cite{HaleviTutorial}. The first generation of HE mainly relies on two hard problems on ideal lattices \cite{ArmGuide}: (i) \emph{Closest Vector Problem} (CVP), where recovering the plaintext is equal to find the closest vector to a point, with respect to a lattice basis; (ii) \emph{Shortest Vector Problem} (SVP), where recovering the plaintext is equal to find the shortest vector in the lattice, with respect to a lattice basis. Although it is a promising approach, a drawback resides in uncontrolled noise growing, requiring mandatory squashing and bootstrapping. The use of these two techniques results in an increased complexity of the scheme and higher computational time. In this phase, efforts are focused on solutions to tackle this issue; examples are \emph{batching} \cite{GH}, which packs multiple plaintexts into a single cyphertext to enable parallel homomorphic evaluation on multiple inputs and removing squashing \cite{FV}. 
In the second generation of HE, the goal is twofold: the containment of noise growth and improved performance. Moreover, the security of these new schemes resides on a more standard hard problem, i.e. (i) \emph{Learning With Errors} (LWE) problem, where recovering the plaintext is equivalent to solve random linear equations, perturbed by noise; or, (ii) \emph{Ring Learning With Errors} (RLWE) problem, which is an enhanced version of LWE in terms of performance. The elimination of squashing represents another essential optimization. The authors in \cite{BV11_2} introduced \emph{modulus switching}, a ``trick'' that reduces noise without knowing the encryption key, making the SWHE bootstrappable. 
Finally, the third generation comprehends further optimizations in terms of performances, as in \cite{GSW}, and improvements in the bootstrapping procedure \cite{Chillotti}, \cite{Micciancio}. The authors in \cite{CKKS} propose support for real and complex numbers. Table \ref{Table_HE_Scheme} summarizes principal results for HE schemes. 
\begin{table*}[]
\caption{SWHE and FHE schemes}
\label{Table_HE_Scheme}
\footnotesize
\centering
\begin{tabular}{|l|l|l|l|} 
\hline
\textbf{HE Generation} & \textbf{Underlying hard problem} & \textbf{HE Scheme} & \textbf{Characteristics} \\ 
\hline
\multirow{4}{*}{\begin{tabular}[c]{@{}l@{}}First: \\problem of noise growth\end{tabular}} & \multirow{3}{*}{Ideal lattices: CVP/SVP} & Gentry \cite{Gentry1} & SSP for squashing \\ 
\cline{3-4}
 &  & GH \cite{GH} & Elimination of squashing~ \\ 
\cline{3-4}
 &  & SV \cite{SV} & Batching \\ 
\cline{2-4}
 & AGCD & DGHV \cite{DGHV} & \begin{tabular}[c]{@{}l@{}}Simplicity\\Over the integer\end{tabular} \\ 
\hline
\multirow{6}{*}{\begin{tabular}[c]{@{}l@{}}Second: \\error control and \\performance improvements\end{tabular}} & \multirow{2}{*}{LWE} & BV2011(a) \cite{BV11_2}~ & Modulus reduction to avoid squashing \\ 
\cline{3-4}
 &  & Bra \cite{Bra} & \begin{tabular}[c]{@{}l@{}}Batching\\Elimination of modulus switching\end{tabular} \\ 
\cline{2-4}
 & \multirow{3}{*}{RLWE} & BV2011(b) \cite{BV11_1} & First scheme in RLWE \\ 
\cline{3-4}
 &  & FV \cite{FV} & Turning of Bra \cite{Bra} into RLWE  \\ 
\cline{3-4}
 &  & BGV12 \cite{BGV12} & \begin{tabular}[c]{@{}l@{}}No bootstrapping \\Batching \\Improvement of modulus switching~\end{tabular} \\ 
\hline
\multirow{4}{*}{\begin{tabular}[c]{@{}l@{}}Third:\\bootstrapping optimization, \\support for approximate arithmetic\end{tabular}} & \multirow{3}{*}{LWE} & \begin{tabular}[c]{@{}l@{}}GSW \cite{GSW} ~\\~\end{tabular} & \begin{tabular}[c]{@{}l@{}}Faster\\Optional bootstrapping~\\No modulus switching\end{tabular} \\ 
\cline{3-4}
 &  & Chillotti \cite{Chillotti} & \begin{tabular}[c]{@{}l@{}}Based on GSW\\Faster bootstrapping (in less than 0.1 sec)\end{tabular} \\ 
\cline{3-4}
 &  & Micciancio \cite{Micciancio}& \begin{tabular}[c]{@{}l@{}}Faster bootstrapping (in less than 1 sec)\end{tabular} \\ 
\cline{2-4}
 & RLWE & CKKS \cite{CKKS} & \begin{tabular}[c]{@{}l@{}}Real/complex number \\Batching\end{tabular} \\
\hline
\end{tabular}
\label{Table_HE}
\end{table*}
\subsection{Available HE Libraries}
Academic and open source communities are currently and extensively working on the implementation of various libraries, making this a dynamic and rapidly changing panorama. Most of the existing libraries are thought for cryptographers and are highly customizable, through the definition of a \emph{context}, namely a container where HE scheme and its parameters can be defined. Typically, existing libraries provide a set of homomorphic operations that can be used to implement complex functions.
Main libraries are:
\begin{itemize}
    \item HElib\footnote{https://github.com/shaih/HElib}, that is in C++ and implements BGV scheme with various optimizations. Support for CKKS is also available;
    \item SEAL\footnote{https://github.com/microsoft/SEAL}, that is in C++ and implements BGV and CKKS; 
    \item PALISADE\footnote{https://gitlab.com/palisade/palisade-release}, that is in C++ and supports BFV, BGV and CKKS;
    \item TFHE\footnote{https://tfhe.github.io/tfhe/}, that implements fast bootstrapping in C/C++ for the HE scheme introduced in \cite{Chillotti}.
\end{itemize}
\section{Privacy-Preserving Data Science: An Overview}
\label{PPDS}
As presented in Section \ref{Motivation}, we are interested in assessing the state-of-the-art solutions for automatic classification, textual analysis, and statistical analysis of confidential data.
The ability to delegate computations to a third party, both in terms of resources -- e.g., cloud --  and competencies -- e.g., data scientists -- fueled improvements in privacy-preserving data science. In the last three years, considerable attention has been devoted to improving the computational time of machine learning tasks over encrypted data, keeping invariant accuracy of results. 
\subsection{Privacy-Preserving Machine Learning} 
\begin{figure*}
    \centering
    \includegraphics[scale=0.50]{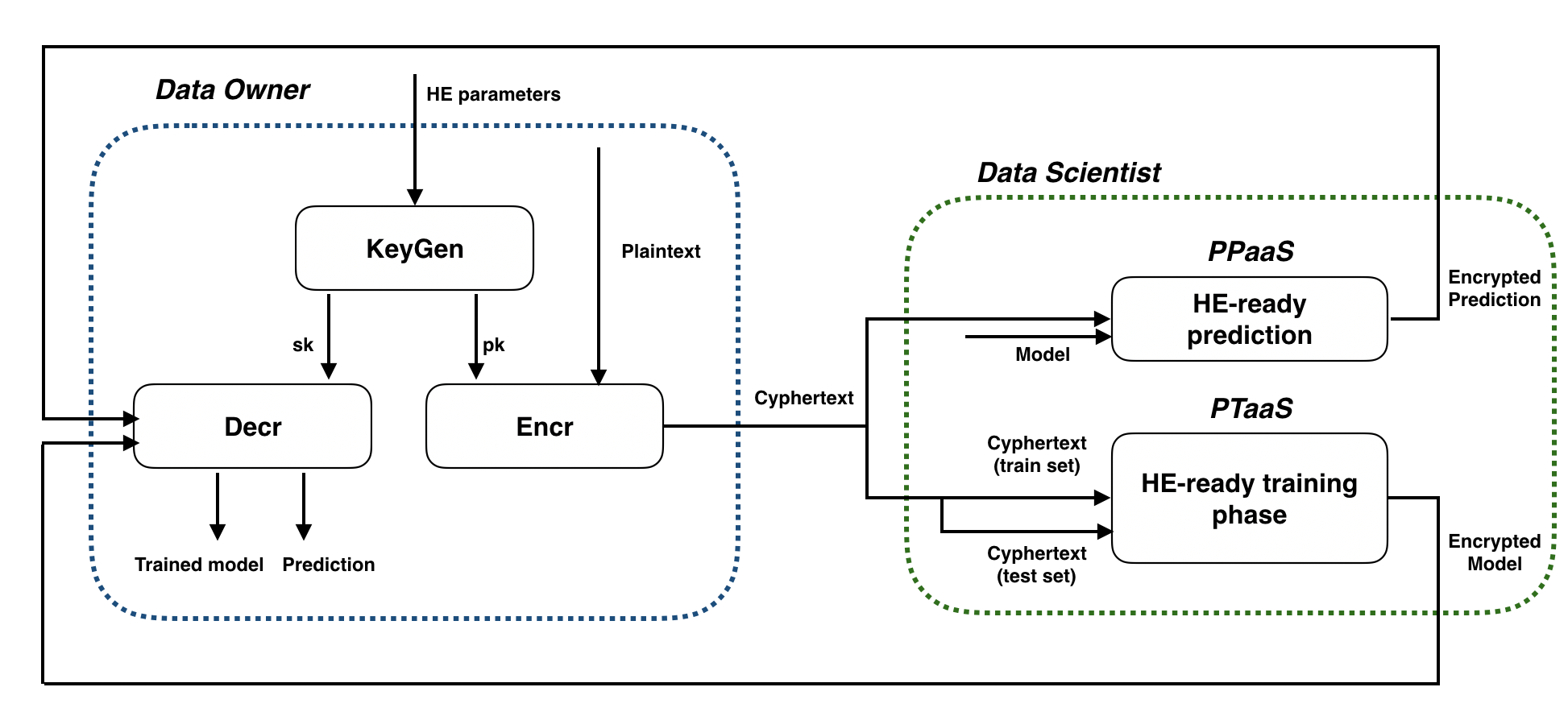}
    \caption{Private Prediction as a Service and Private Training as a Service}
    \label{fig:PPaaSPTaaS}
\end{figure*}

As in Fig. \ref{fig:PPaaSPTaaS}, two scenarios are explored. The first one is \emph{Private Prediction as a Service} (PPaaS), where a data owner outsources prediction $P(\Psi)$ of private encrypted data $\Psi$ to a third party, e.g. an untrusted data scientist, that has an already trained model $w$ and, potentially, computational resources to perform the prediction task. The second one is \emph{Private Training as a Service} (PTaaS), where a data owner supplies encrypted data $\Psi$ to a third party for training a HE-ready model $w_{encr}$ and, then, use $w_{encr}$ to perform prediction on fresh ciphertexts; the obtained model is ready to use for prediction tasks. In both scenarios, the encrypted results are then sent back to the data owner to decrypt them. In the first scenario, the data owner should learn only the result of prediction $P_{encr}(\Psi)$, but nothing about the model $w$ that may represent a private asset for the untrusted data scientist. This model is trained on plaintext data in the availability of data scientist. 

For the PPaaS scenario, predictions are mainly obtained via: (i) a logistic regression, as in \cite{HEPipeline}, \cite{Cheon1}, \cite{BosMedical}; (ii) Neural Networks (NN), as in \cite{Cryptonets}, \cite{Lola}, \cite{CryptoDL}, \cite{Chabanne}.
Progresses on GPU adoption for prediction are also present: the authors in \cite{CNNGPU1}  use GPU to improve computational time while running Convolutional Neural Networks (CNN) on encrypted data, while in PrivFT \cite{PrivFT}, the authors provide a GPU version of CKKS scheme. Bost et al. \cite{BostML} implement and train three classifiers: hyperplane decision, N\"aive Bayes and decision trees.
For what concerns PTaaS scenario, most works explore training of logistic regression, as in \cite{GentryLogRegr}, \cite{LogRegr30k}, \cite{Cheon1}, \cite{Kim1}. In 2019, Nandakumar et al. \cite{Nandakumar} showed that training a NN over encrypted data is viable. 

Extensive use of LHE or SWHE is done, supported by HElib in most cases, as in \cite{Cryptonets} or in \cite{Lola}, where the authors use a BFV scheme. As seen in Section \ref{SWHE_FHE}, LHE allows computation of fixed low degree polynomials, representing a limit when required to deal with complex functions, as the HE parameters grow. Furthermore, the additional computational time as a result of bootstrapping may represent a bottleneck for real-world applications. Gentry et al. in \cite{GentryLogRegr} showed that, through various optimizations, it is possible to use an FHE scheme during the training of logistic regression, based on BGV scheme, while in \cite{Nandakumar} an approach for using FHE during the training of neural networks is presented. The use of parallelization also enhances the process of bootstrapping, as in \cite{Cheon1}.
Choosing LHE or SWHE helps in coping with non-polynomial functions. For learning tasks, we need activation and loss functions, which are not \emph{homomorphic-ready}. 
We remember that the set of the admitted circuits to be evaluated by the HE scheme should be expressed by $AND$ and $XOR$: addition and multiplication are supported, i.e., only polynomial functions can be evaluated efficiently.
Rectified Linear Unit (ReLU), sigmoid, or Tanh should be approximated as polynomial, and, for the sake of noise containment, these polynomials should be of low degree.
One can resort at the Taylor series approximation of these functions \cite{Chabanne}, but, as showed in \cite{Cheon1}, this approximation is accurate only around a fixed point. Other methods available in literature are: (i) look-up table, as in \cite{Nandakumar}; (ii) (standard or modified) Chebychev polynomial, as in \cite{CryptoDL}; (iii) least squares as in \cite{Kim2}.

Moreover, the same difficulties arise when encountering other operations such as comparison, sorting, exponentials, logarithms, division, pooling, etc. A viable approach is to construct a set of building blocks implementing homomorphic versions of non-polynomial operations as in \cite{GentryLogRegr} and in \cite{BostML}. These building modules are common to different algorithms and constitute a toolkit to ease the implementation of learning tasks.

In the area of natural language processing, appealing results are obtained in (i) PrivFT \cite{PrivFT}, where the authors focus on text classification and training of a shallow NN, \emph{fasttext}; (ii)\cite{CryptoEmbedd}, where the authors present sentiment classification of IMDb dataset using Recurrent Neural Network (RNN); (iii) \cite{Costantino}, where the authors use a bag-of-word approach for the binary classification problem of detection of terrorist Twitter accounts. 

In Table \ref{PPaaS Table} and in Table \ref{PPaaT_Table}, we summarize main approaches for PPaaS and for PTaaS respectively. Since the obtained accuracy seems acceptable for most real-world applications, we focus on giving results for computational time and employed datasets to compare different contributions. 
\subsection{Statistical Analysis}
\label{Stats}
For statistical analysis, we intend classical estimations of mean, variance, histogram, linear regression etc. on a large scale dataset with a great number of dimensions \cite{FHEStats}.
Employing homomorphic encryption for statistical analysis may improve the accuracy of metrics with respect to the application of randomization, but analogous issues as the one encountered for privacy-preserving machine learning arise. In \cite{SecureStats}, to avoid the computation of division, the authors first performs the needed summation of encrypted data, as follows: 
\begin{equation}
\sum^{N-1}_{i=0}x_i,~\sum^{N-1}_{i=0}y_i,~\sum^{N-1}_{i=0}y^2_i,\sum^{N-1}_{i=0}x^2_i,~\sum^{N-1}_{i=0}x_iy_i
\end{equation}
where $(X,Y) = (x_0,y_0,) \ldots, (x_{N-1},y_{N-1})$ are $N$ independent pair of integers.

These values act as building blocks for mean and covariance and are sent to the data owner, which, after decryption, will compute the required division to obtain the metrics. Analogously, the authors in \cite{FHEStats} construct primitives for matrix and ``greater than" operations, providing tools to compute histogram and k-percentile for categorical data, and principal component analysis and linear regression for numerical data. In \cite{LargeScaleStats}, the authors apply matrix-matrix product to implement linear regression. Computational time is feasible for most real-world applications, e.g. in \cite{FHEStats}, training a regression model with about $30k$ samples of data takes under $20$ minutes.
%
%
%
\begin{table*}[ht!]
\label{PPaaS Table}
\caption{Private Prediction as a Service solutions}
\centering
\resizebox{\linewidth}{!}{%
\begin{tabular}{|l|l|l|l|l|l|} 
\hline
\textbf{Reference} & \textbf{PPaaS Model} & \textbf{HE scheme} & \textbf{Platform} & \textbf{Running Time} & \textbf{Dataset} \\ 
\hline
\multirow{2}{*}{Graepel et al. \cite{MLconfidential}} & Linear Means Classifier & \multirow{2}{*}{FV / Bra} & \multirow{2}{*}{\begin{tabular}[c]{@{}l@{}}Intel Core i7 \\@2.8 GHz with 8GB of RAM\end{tabular}} & 6 sec & \multirow{2}{*}{\begin{tabular}[c]{@{}l@{}}Wisconsin Breast \\Cancer Data\end{tabular}} \\ 
\cline{2-2}\cline{5-5}
 & \begin{tabular}[c]{@{}l@{}}Fisher’s Linear \\Discriminant classifier\end{tabular} &  &  & 20 sec &  \\ 
\hline
Gilad-Bachrac et al.\cite{Cryptonets} & Prediction by NN with 5 layers & \begin{tabular}[c]{@{}l@{}}HE scheme based on \cite{Lopez}\\SEAL\end{tabular} & \begin{tabular}[c]{@{}l@{}}Intel Xeon E5-1620\\@ 3.5GHz with 16GB RAM\end{tabular} & 570 sec for a prediction & MNIST \\ 
\hline
Costantino et al. \cite{Costantino} & Bag-of-word & BGV on HElib & \begin{tabular}[c]{@{}l@{}}Intel Core i7-6700 \\@ 3.40GHz with GB RAM\end{tabular} & 19 min & Tweets \\ 
\hline
\multirow{2}{*}{Hesamifard et al. \cite{CryptoDL}} & \multirow{2}{*}{\begin{tabular}[c]{@{}l@{}}Prediction by Convolutional \\Neural Network with 6 layers\end{tabular}} & \multirow{2}{*}{LHE HElib} & \multirow{2}{*}{\begin{tabular}[c]{@{}l@{}}Intel Xeon E5-2640 \\@2.4GHz with 16GB RAM\end{tabular}} & 320 sec & MNIST \\ 
\cline{5-6}
 &  &  &  & 11686 sec & CIFAR-10 \\ 
\hline
Masters et al. \cite{HEPipeline}& \begin{tabular}[c]{@{}l@{}}Nesterov’s Accelerate \\GD-based logistic regression\end{tabular} & CKKS HElib & Titan V & 4500 speed up on mult & \begin{tabular}[c]{@{}l@{}}Banco Bradesco \\financial transactions\end{tabular} \\ 
\hline
\multirow{2}{*}{Brutzkus et al. \cite{Lola}} & \begin{tabular}[c]{@{}l@{}}Same NN as CryptoNets \\for prediction\end{tabular} & \multirow{2}{*}{BFV} & \multirow{2}{*}{\begin{tabular}[c]{@{}l@{}}Azure standard VM \\with 8 vCPUs 32GB of RAM\end{tabular}} & 0.29 sec for a prediction & MNIST \\ 
\cline{2-2}\cline{5-6}
 & \begin{tabular}[c]{@{}l@{}}Linear model trained with\\features generated by AlexNet\end{tabular} &  &  & 0.16 sec for prediction & CalTech-101 \\ 
\hline
Al Badawi et al. \cite{PrivFT} & \begin{tabular}[c]{@{}l@{}}Fasttext NN trained \\on plaintext data\end{tabular} & CKKS on GPU & \begin{tabular}[c]{@{}l@{}}NVIDIA DGX-1 multi-GPU\\with 8 V100 cards\end{tabular} & 0.66 sec & AGNews \\ 
\hline
Podschwadt et al. \cite{CryptoEmbedd} & \begin{tabular}[c]{@{}l@{}}Embedding layer + \\RNN layer with 128 units\end{tabular} & CKKS on HElib & \begin{tabular}[c]{@{}l@{}}AMD Ryzen 5 2600\\@ 3.5GHz with 32GB RAM.\end{tabular} & \begin{tabular}[c]{@{}l@{}}547.6 sec for a batch \\of 128 samples\end{tabular} & IMDb \\
\hline
\end{tabular}
}
\end{table*}
\subsection{Private DS-ready Libraries} \label{PPDS Libraries}
Recently, academia and open source communities accepted the challenge to make secure computation through HE appealing and affordable to a non expert audience had been performed. The necessity to let HE more accessible gave rise to frameworks that are specifically devoted to data scientists, and consequently are less customizable, masking almost all cryptographic choice of parameters. 

The development of Python HE libraries can reach pure data scientists with little cryptography background and provide a friendly and well-known interface and API.
OpenMined\footnote{\url{https://www.openmined.org}}, proposed a library called \emph{PySyft} \cite{Trask}, which is based on PyTorch. It enables privacy-preserving deep learning, using HE for single data or model owner.
\emph{PySyft} already provides support for the Paillier scheme, and implementation of FV scheme is in progress. 
OpenMined has released a tool for Private NLP: \emph{SyferText} \cite{SyferText} enables creation and training of deep learning NLP model over the local and distributed confidential dataset and the encapsulation of the model in an NLP pipeline. The library is built on top of PySyft and provides the ability to pre-process and encrypt text data.
Cape Privacy\footnote{\url{https://capeprivacy.com}} developed \emph{TF Encrypyted} \cite{TFEncrypted}, a framework built on top of TensorFlow. The user is allowed to inspect the static data flow graph of its pipeline by \emph{TensorBoard}, which helps discover and manage machine learning and cryptography issues. Support for Keras is one of the most appealing issues in progress.
However, these libraries are still at the research level.
%
%
%
%
\begin{table*}[ht!]
\label{PPaaT_Table}
\caption{Private Training as a Service solutions}
\centering
\resizebox{\linewidth}{!}{%
\begin{tabular}{|l|l|l|l|l|l|} 
\hline
\textbf{Task} & \textbf{Reference} & \textbf{HE scheme} & \textbf{Platform} & \textbf{Running Time} & \textbf{Dataset} \\ 
\hline
\multirow{6}{*}{\begin{tabular}[c]{@{}l@{}}PTaaS - \\Logistic Regression\end{tabular}} & Gentry et al. \cite{GentryLogRegr} & \begin{tabular}[c]{@{}l@{}}BGV with\\bootstrapping onHElib\end{tabular} & \begin{tabular}[c]{@{}l@{}}Intel Xeon E5-2698 v3 \\@2.30GHz with 250GB RAM\end{tabular} & \begin{tabular}[c]{@{}l@{}}More than 4 hours; \\one hour if \\multithreading is used\end{tabular} & \begin{tabular}[c]{@{}l@{}}iDASH competition \\data \end{tabular} \\ 
\cline{2-6}
 & Kim et al. \cite{Kim1} & CKKS & \begin{tabular}[c]{@{}l@{}}Intel Xeon CPU E5-2620 \\@2.10 GHz\end{tabular} & 6 mins & \begin{tabular}[c]{@{}l@{}}iDASH competition\\~data 2017 \end{tabular} \\ 
\cline{2-6}
 & Chen at al. \cite{Chen} & FV on SEAL & \begin{tabular}[c]{@{}l@{}}Intel(R) Xeon(R) CPU E3-1280 \\@ 3.70GHz with 16GB RAM\end{tabular} & \begin{tabular}[c]{@{}l@{}}3.2 hours to 0.4 hours \\for 1-bit GD\end{tabular} & \begin{tabular}[c]{@{}l@{}}iDASH competition\\~data 2017 \end{tabular} \\ 
\cline{2-6}
 & Bergamaschi et al. \cite{LogRegr30k} & CKKS on HElib & \begin{tabular}[c]{@{}l@{}}Intel E5-2640 \\@2.5GHz, with 64 GB RAM\end{tabular} & 20 min & \begin{tabular}[c]{@{}l@{}}iDASH competition\\~data 2018 \end{tabular} \\ 
\cline{2-6}
 & \multirow{2}{*}{Han et al. \cite{Cheon1}} & \multirow{2}{*}{CKKS} & \multirow{2}{*}{\begin{tabular}[c]{@{}l@{}}IBM POWER8 \\@ 4.0GHz with 256GB RAM\end{tabular}} & 17 hours & \begin{tabular}[c]{@{}l@{}}Korean credit bureau \end{tabular} \\ 
\cline{5-6}
 &  &  &  & 2 hours & MNIST \\ 
\hline
\multirow{3}{*}{\begin{tabular}[c]{@{}l@{}}PTaaS -\\Neural Network\end{tabular}} & Nandakumar et al. \cite{Nandakumar} & HElib & \begin{tabular}[c]{@{}l@{}}Intel Xeon E5-2698 v3 \\@2.30GHz with 250GB RAM\end{tabular} & \begin{tabular}[c]{@{}l@{}}1.5 days for NN1 (954 nodes, \\50 epochs)\\40 min for NN2 (122 nodes, \\50 epochs)\end{tabular} & \begin{tabular}[c]{@{}l@{}}MNIST \\(mini-batch of 60 training samples)\end{tabular} \\ 
\cline{2-6}
 & \multirow{2}{*}{Al Badawi et al. \cite{PrivFT}} & CKKS on SEAL & 104 CPU cores & 11.1 days (fasttext NN) & \multirow{2}{*}{\begin{tabular}[c]{@{}l@{}}YouTube\\spam collection\end{tabular}} \\ 
\cline{3-5}
 &  & CKKS on GPU & \begin{tabular}[c]{@{}l@{}}NVIDIA DGX-1 multi-GPU \\with 8 V100 cards\end{tabular} & 5.04 days(fasttext NN) &  \\
\hline
\end{tabular}
}
\end{table*}
\section{Discussion and Challenges} \label{OpenIssues}
In this section, we discuss the application of HE schemes in data science applications for an enterprise context.

The main advantage of HE is not only the privacy of data but also the so-called \emph{circuit privacy} \cite{HaleviTutorial}; it ensures that no information about $C^*$ (and the related function $f$) is leaked by the output, even if public and secret keys are available. This kind of protection assures to the data scientist that considers his model a valuable asset, constituting a reliable guarantee when \emph{obfuscation} of the algorithm \cite{ArmGuide} cannot be employed. Another strength of HE is non-interactivity \cite{CryptoDL}: the data owner sends encrypted data to the data scientist, which homomorphically executes classification, training, or statistical analysis of data. Besides this exchange, the data owner is released from the computation phase; instead, in other approaches, such as Multi-Party Computation, the data owner is required to be online for all the computational time. 
However, HE does not come without drawbacks: besides the well-known computational overhead, we have to consider two other aspects. One is the lack of the support of multiple users for most popular HE schemes \cite{HaleviTutorial}: in almost all scenarios, we suppose that all input data are encrypted under the same key. Let us consider the scenario in which multiple data owners, who may not trust each other, are willing to train a model on their joint encrypted data. This scenario can be enabled by resorting to \emph{Multikey} homomorphic encryption that ensures data encryption under multiple unrelated keys and in our considered scenario would allow the recovery of the trained model from the different data owners, each with their key \cite{Lopez}. Another interesting issue is the lack of verifiability of the computation, also called “integrity” in \cite{HaleviTutorial}: the data owner cannot verify if the result of the computation is correct.  

However, as assessed in Section \ref{PPDS}, it emerges that HE is becoming able to be employed in real-world data science applications. The reported computational time for classification and even training on encrypted data seems promising, and the recent advances on GPU support for HE libraries are encouraging; the same holds for statistical analysis as described in Section \ref{Stats}.
Notable examples of employment of HE for prediction on real-world data are in \cite{Cheon1}, \cite{BosMedical}, \cite{HEPipeline}. The authors in \cite{Cheon1} trained a logistic regression on the financial dataset from the Korea Credit Bureau to assess clients' credit ratings for a given threshold. For a dataset of more than 400k and 200 features, the training took 17 hours.
In \cite{BosMedical}, a cloud working implementation of the prediction of cardiovascular disease is presented; moreover, they propose a module for automatic parameter selection of the chosen LHE. This phase is critical because parameters influence the security and compactness of the scheme and the performance, which depends on the complexity of the function to be evaluated.
The HE Standardardization group \cite{HEStandard} also tackles the problem of the choice of parameters, and tables of recommended parameters are given to guide the data scientist in the choice, having fixed the desired security level.

The integration of HE into already existing data science applications could be hard to accomplish, so an entirely novel approach must be developed. The authors in \cite{HEPipeline} proposed a homomorphic machine learning pipeline for the prediction of financial data (more than 36K entries) of the Brazilian Banco Bradesco. In the pipeline, there are trusted containers involved during key generation, encryption of confidential data, and decryption of evaluated encrypted data, while computation happens in an untrusted container, as expected. Data collection and preparation are performed as in plain machine learning pipeline. They also propose an approach to feature engineering on encrypted data \cite{HEPipeline}. 

Another issue is the unavailability of support tools for developers \cite{GentryLogRegr}: HE parameters' choice, data representation, and circuit design are all left to the developer. In \cite{HEAPI}, the goal is to provide an API that translates business logic by data scientist into a low-level language, the \emph{Assembly Language} of HE, while in \cite{TFEncrypted} Tensorboard constitutes a tool for manage the execution of the privacy-preserving computation. 
It emerges that, though privacy-preserving data science with HE is still at the pilot stage, the available results assure that for our purposes could become a promising approach. We expect that the development of tools such as the ones explored in Section \ref{PPDS Libraries} will facilitate the use of HE in data science applications.  

\section{Conclusions}
In this paper, we provide an overview of the state-of-the-art in privacy-preserving data science with HE. We also describe the use cases of potential interest for the Bank of Italy. Finally, we discuss challenges and open points concerning the development of real-world applications. HE is a promising approach for enterprise applications to allow collaboration between data owners and untrusted data scientists and researchers. As this is a fertile research area, we expect that new developments, especially in support of the data scientist, will also enable its use in a production context. 
\section*{Acknowledgments}
The author would like to thank Dr. Marco Benedetti and the Applied Research Team (IT Department, Bank of Italy) for support. %

\end{document}